\documentstyle[12pt]{article}

\advance \topmargin by -\headheight
\advance \topmargin by -\headsep
     
\evensidemargin \oddsidemargin
 \date{}    
\begin{document}
\newcommand{\sect}[1]{\setcounter{equation}{0}\section{#1}}
\renewcommand{\theequation}{\thesection.\arabic{equation}}

\topmargin -.6in
\def\nonu{\nonumber}
\def\rf#1{(\ref{eq:#1})}
\def\lab#1{\label{eq:#1}} 
\def\br{\begin{eqnarray}}
\def\er{\end{eqnarray}}
\def\be{\begin{equation}}
\def\ee{\end{equation}}
\def\0{\nonumber}
\def\lb{\lbrack}
\def\rb{\rbrack}
\def\({\left(}
\def\){\right)}
\def\v{\vert}
\def\bv{\bigm\vert}
\def\lskip{\vskip\baselineskip\vskip-\parskip\noindent}
\relax
\newcommand{\nit}{\noindent}
\newcommand{\ct}[1]{\cite{#1}}
\newcommand{\bi}[1]{\bibitem{#1}}
\def\a{\alpha}
\def\b{\beta}
\def\ca{{\cal A}}
\def\cm{{\cal M}}
\def\cn{{\cal N}}
\def\cf{{\cal F}}
\def\d{\delta} 
\def\D{\Delta}
\def\eps{\epsilon}
\def\g{\gamma}
\def\G{\Gamma}
\def\grad{\nabla}
\def\h{ {1\over 2}  }
\def\hc{\hat{c}}
\def\hd{\hat{d}}
\def\hg{\hat{g}}
\def\hp{ {+{1\over 2}}  }
\def\hm{ {-{1\over 2}}  }
\def\k{\kappa}
\def\l{\lambda}
\def\L{\Lambda}
\def\lg{\langle}
\def\m{\mu}
\def\n{\nu}
\def\o{\over}
\def\om{\omega}
\def\O{\Omega}
\def\p{\phi}
\def\pa{\partial}
\def\pr{\prime}
\def\ra{\rightarrow}
\def\rh{\rho}
\def\rg{\rangle}
\def\s{\sigma}
\def\t{\tau}
\def\th{\theta}
\def\ti{\tilde}
\def\wti{\widetilde}
\def\inte{\int dx }
\def\xb{\bar{x}}
\def\yb{\bar{y}}

\def\tr{\mathop{\rm tr}}
\def\Tr{\mathop{\rm Tr}}
\def\partder#1#2{{\partial #1\over\partial #2}}
\def\ds{{\cal D}_s}
\def\wtwo{{\wti W}_2}
\def\lie{{\cal G}}
\def\alie{{\widehat \lie}}
\def\dlie{{\cal G}^{\ast}}
\def\elie{{\widetilde \lie}}
\def\edlie{{\elie}^{\ast}}
\def\hlie{{\cal H}}
\def\wlie{{\widetilde \lie}}

\def\rlx{\relax\leavevmode}
\def\inbar{\vrule height1.5ex width.4pt depth0pt}
\def\IZ{\rlx\hbox{\sf Z\kern-.4em Z}}
\def\IR{\rlx\hbox{\rm I\kern-.18em R}}
\def\IC{\rlx\hbox{\,$\inbar\kern-.3em{\rm C}$}}
\def\one{\hbox{{1}\kern-.25em\hbox{l}}}

\def\PRL#1#2#3{{\sl Phys. Rev. Lett.} {\bf#1} (#2) #3}
\def\NPB#1#2#3{{\sl Nucl. Phys.} {\bf B#1} (#2) #3}
\def\NPBFS#1#2#3#4{{\sl Nucl. Phys.} {\bf B#2} [FS#1] (#3) #4}
\def\CMP#1#2#3{{\sl Commun. Math. Phys.} {\bf #1} (#2) #3}
\def\PRD#1#2#3{{\sl Phys. Rev.} {\bf D#1} (#2) #3}
\def\PLA#1#2#3{{\sl Phys. Lett.} {\bf #1A} (#2) #3}
\def\PLB#1#2#3{{\sl Phys. Lett.} {\bf #1B} (#2) #3}
\def\JMP#1#2#3{{\sl J. Math. Phys.} {\bf #1} (#2) #3}
\def\PTP#1#2#3{{\sl Prog. Theor. Phys.} {\bf #1} (#2) #3}
\def\SPTP#1#2#3{{\sl Suppl. Prog. Theor. Phys.} {\bf #1} (#2) #3}
\def\AoP#1#2#3{{\sl Ann. of Phys.} {\bf #1} (#2) #3}
\def\PNAS#1#2#3{{\sl Proc. Natl. Acad. Sci. USA} {\bf #1} (#2) #3}
\def\RMP#1#2#3{{\sl Rev. Mod. Phys.} {\bf #1} (#2) #3}
\def\PR#1#2#3{{\sl Phys. Reports} {\bf #1} (#2) #3}
\def\AoM#1#2#3{{\sl Ann. of Math.} {\bf #1} (#2) #3}
\def\UMN#1#2#3{{\sl Usp. Mat. Nauk} {\bf #1} (#2) #3}
\def\FAP#1#2#3{{\sl Funkt. Anal. Prilozheniya} {\bf #1} (#2) #3}
\def\FAaIA#1#2#3{{\sl Functional Analysis and Its Application} {\bf #1} (#2)
#3}
\def\BAMS#1#2#3{{\sl Bull. Am. Math. Soc.} {\bf #1} (#2) #3}
\def\TAMS#1#2#3{{\sl Trans. Am. Math. Soc.} {\bf #1} (#2) #3}
\def\InvM#1#2#3{{\sl Invent. Math.} {\bf #1} (#2) #3}
\def\LMP#1#2#3{{\sl Letters in Math. Phys.} {\bf #1} (#2) #3}
\def\IJMPA#1#2#3{{\sl Int. J. Mod. Phys.} {\bf A#1} (#2) #3}
\def\AdM#1#2#3{{\sl Advances in Math.} {\bf #1} (#2) #3}
\def\RMaP#1#2#3{{\sl Reports on Math. Phys.} {\bf #1} (#2) #3}
\def\IJM#1#2#3{{\sl Ill. J. Math.} {\bf #1} (#2) #3}
\def\APP#1#2#3{{\sl Acta Phys. Polon.} {\bf #1} (#2) #3}
\def\TMP#1#2#3{{\sl Theor. Mat. Phys.} {\bf #1} (#2) #3}
\def\JPA#1#2#3{{\sl J. Physics} {\bf A#1} (#2) #3}
\def\JSM#1#2#3{{\sl J. Soviet Math.} {\bf #1} (#2) #3}
\def\MPLA#1#2#3{{\sl Mod. Phys. Lett.} {\bf A#1} (#2) #3}
\def\JETP#1#2#3{{\sl Sov. Phys. JETP} {\bf #1} (#2) #3}
\def\JETPL#1#2#3{{\sl  Sov. Phys. JETP Lett.} {\bf #1} (#2) #3}
\def\PHSA#1#2#3{{\sl Physica} {\bf A#1} (#2) #3}
\def\PHSD#1#2#3{{\sl Physica} {\bf D#1} (#2) #3}

\begin{center}
{\large\bf  Classical Integrable Super  sinh-Gordon equation with  Defects}
\end{center}
\normalsize
\vskip .4in

\begin{center}
J.F. Gomes,  L.H. Ymai and A.H. Zimerman

\par \vskip .1in \noindent
Instituto de F\'{\i}sica Te\'{o}rica-UNESP\\
Rua Pamplona 145\\
01405-900 S\~{a}o Paulo, Brazil
\par \vskip .3in

\end{center}

\begin{abstract}
 
The introduction of defects  is  discussed under the Lagrangian formalism  and 
Backlund transformations for the $N=1$ super sinh-Gordon model.   
Modified conserved momentum and energy are constructed for this case. 
Some explicit examples of different Backlund solitons  solutions  are discussed.   
The Lax formulation within the  space split by the defect leads to the integrability 
of the model and henceforth to the existence of an
infinite number of constants of motion.

\end{abstract}

\section{Introduction}
A quantum integrable theory of defects 
involving  free bosonic and free fermionic fields  was first studied in ref.  \cite{mussardo} 
 following the achievements obtained in studying
the quantum field theory with boundaries \cite{fring}, \cite{zam}.
The Lagrangian formulation of a class of relativistic integrable field theories 
admiting certain discontinuities has been 
studied recently  \cite{bowcock1} -  \cite{corrigan}.
In particular, in ref. \cite{bowcock1} the authors  have considered  a field theory in which different soliton 
solutions of the sine-Gordon model  are linked in such a way that the integrability is preserved.
The integrability of the total system imposes severe constraints specifying  the possible
types of defects.  These are characterized by Backlund transformations which are known to  
connect two different soliton solutions.

The supersymmetric $N=1$ sinh-Gordon in terms of superfields was  proposed in ref. \cite{chai}  by introducing a pair of 
Grassmann coordinates.  The corresponding Backlund transformation was proposed  in terms of superfields also.
An important piece of information characterizing the defect is given by boundary functions which are consistent 
with the Backlund transformations and  leads to the construction of modified conserved momentum and energy.
 The aim of  this paper is to  consider a classical  interacting field  theory containing both Bose  and Fermi fields.  
We generalize the Lagrangian approach of \cite{bowcock1} -  \cite{corrigan} to include the 
$N=1$ supersymmetric sinh-Gordon in the presence of integrable defects described by Backlund transformation.
We explicitly construct the boundary functions consistently with the
Backlund transformations.  Several cases of transition through the defect are studied. 
In particular, when the fermionic fields vanish, the pure bosonic soliton case 
of refs. \cite{bowcock1} - \cite{corrigan} is recovered. On the other hand, for vanishing bosonic 
fields, we find the pure fermionic case.  More interesting cases occur when we 
consider transitions between vacuum  to a system of Bose and
Fermi fields and transitions between two distinct configurations of 
non trivial Bose and Fermi states.  The integrability of the system is ensured by 
the zero curvature representation of the equations of motion.

This paper is organized as follows. 
In Sect. 2 we discuss the Lagrangian approach  to describe  these integrable 
supersymmetric models with defects. In particular, specific boundary terms  are
chosen in order to ensure modified energy momentum conservation.  Explicit examples 
of pure free bosonic and free fermionic fields together with the
supersymmetric sinh-Gordon are also discussed. Various  Backlund solutions for the 
super sinh-Gordon (sine-Gordon) models are discussed in  Sect.3. 
In Sect. 4 we present the zero curvature  representation  of the supersymmetric sinh-Gordon model
equations of motion. By introducing two regions  around 
the defect \cite{bowcock1} we explicitly construct, in a closed form, a gauge group 
element connecting the Lax pair in the overlap region.  
This fact guarantees the existence of an infinite  set of conservation laws.
In the appendix we  present, in components the Backlund transformation for the super sinh-Gordon model.


\section{Integrable Supersymmetric $sl(2,1)$ Field Theory with a Defect}

\subsection{Lagrangian Description}

The starting point is the Lagrangian density describing bosonic, $\phi_1$ and fermionic, $\bar \psi_1, \psi_1 $ fields
 in the region $x<0$
and correspondingly $\phi_2$ and $\bar \psi_2, \psi_2 $
 in the region $x>0$.  
 A defect contribution located at $x=0$ can be  introduced such that the model under study 
 is described by the Lagrangian density
\begin{eqnarray} 
{\mathcal{L}}=\theta(-x){\mathcal{L}}_{1}+\theta(x){\mathcal{L}}_{2}+\delta(x){\mathcal{L}}_{D},
\label{lll}
\end{eqnarray}
where
\begin{eqnarray}
{\mathcal{L}}_{p}&=&\frac{1}{2}(\partial_{x}\phi_{p})^{2}-\frac{1}{2}(\partial_{t}\phi_{p})^{2}-
\bar{\psi}_{p}\partial_{x}\bar{\psi}_{p}+\bar{\psi}_{p}\partial_{t}\bar{\psi}_{p}+
\psi_{p}\partial_{x}\psi_{p}+\psi_{p}\partial_{t}\psi_{p}\nonumber\\
&&+V_{p}(\phi_{p})+W_{p}(\phi_{p},\psi_{p},\bar{\psi}_{p}), \qquad  p=1,2 \nonumber\\
{\mathcal{L}}_{D}&=&\frac{1}{2}\left(\phi_{2}\partial_{t}\phi_{1}-\phi_{1}\partial_{t}\phi_{2}\right)-
\psi_{1}\psi_{2}-\bar{\psi}_{1}\bar{\psi}_{2}+2f_{1}\partial_{t}f_{1}\nonumber\\
&& +B_{0}(\phi_{1},\phi_{2})+B_{1}(\phi_{1},\phi_{2},\psi_{1},\psi_{2},\bar{\psi}_{1},\bar{\psi}_{2},f_{1}),
\end{eqnarray}
 $f_1$ is an auxiliary fermionic field and $V_p, W_p$ correspond to field potentials.
 The quantities $ B_{0}$ and $ B_{1}$ are boundary functions  describing the defect. 
The field equations in the two regions together with the defect conditions at $x=0$ are then given by
\begin{eqnarray}\label{eqmov1}
\partial_{x}^{2}\phi_{1}-\partial_{t}^{2}\phi_{1}&=&\partial_{\phi_{1}}V_{1}+\partial_{\phi_{1}}W_{1},\nonumber\\
\partial_{x}\psi_{1}+\partial_{t}\psi_{1}&=&-\frac{1}{2}\partial_{\psi_{1}}W_{1},\nonumber\\
-\partial_{x}\bar{\psi}_{1}+\partial_{t}\bar{\psi}_{1}&=&-\frac{1}{2}\partial_{\bar{\psi}_{1}}W_{1},\quad \quad x<0
\end{eqnarray}
and 
\begin{eqnarray}\label{eqmov2}
\partial_{x}^{2}\phi_{2}-\partial_{t}^{2}\phi_{2}&=&\partial_{\phi_{2}}V_{2}+\partial_{\phi_{2}}W_{2},\nonumber\\
\partial_{x}\psi_{2}+\partial_{t}\psi_{2}&=&-\frac{1}{2}\partial_{\psi_{2}}W_{2},\nonumber\\
-\partial_{x}\bar{\psi}_{2}+\partial_{t}\bar{\psi}_{2}&=&-\frac{1}{2}\partial_{\bar{\psi}_{2}}W_{2},\quad \quad x>0
\end{eqnarray}
For $x=0$ we find 
\begin{eqnarray}
\partial_{x}\phi_{1}-\partial_{t}\phi_{2}&=&-\partial_{\phi_{1}}B_{0}-\partial_{\phi_{1}}B_{1},\label{df1}\\
\partial_{x}\phi_{2}-\partial_{t}\phi_{1}&=&\partial_{\phi_{2}}B_{0}+\partial_{\phi_{2}}B_{1},\label{df2}\\
\psi_{1}+\psi_{2}&=&\partial_{\psi_{1}}B_{1}=-\partial_{\psi_{2}}B_{1},\label{df3}\\
\bar{\psi}_{1}-\bar{\psi}_{2}&=&-\partial_{\bar{\psi}_{1}}B_{1}=-\partial_{\bar{\psi}_{2}}B_{1},\label{df4}\\
\partial_{t}f_{1}&=&-\frac{1}{4}\partial_{f_{1}}B_{1}.\label{df5}
\end{eqnarray}
with $\partial_{\psi_{p}}=\frac{\partial}{\partial{\psi_{p}}}$ are  fermionic derivatives
acting on the left (the same holds for $\partial_{\bar{\psi}_{p}}$ and $\partial_{f_{1}}$).

Let us first 
consider the time derivative of the momentum:
\begin{eqnarray}
\frac{dP}{dt}&=&\frac{d}{dt}\left[\int_{-\infty}^{0}dx\left(\partial_{t}\phi_{1}\partial_{x}\phi_{1}-
\bar{\psi}_{1}\partial_{x}\bar{\psi}_{1}-\psi_{1}\partial_{x}\psi_{1}\right)\right.\nonumber\\
&&\left.+\int_{0}^{+\infty}dx\left(\partial_{t}\phi_{2}\partial_{x}\phi_{2}-
\bar{\psi}_{2}\partial_{x}\bar{\psi}_{2}-\psi_{2}\partial_{x}\psi_{2}\right)\right].
\end{eqnarray}
From the equations (\ref{eqmov1}), (\ref{eqmov2}) and ignoring contributions from $\pm\infty$, we can write: 
\begin{eqnarray}\label{momento1}
\frac{dP}{dt}&=&\left[\frac{1}{2}(\partial_{x}\phi_{1})^{2}+\frac{1}{2}(\partial_{t}\phi_{1})^{2}-
\bar{\psi}_{1}\partial_{t}\bar{\psi}_{1}-\psi_{1}\partial_{t}\psi_{1}-V_{1}-W_{1}\right.\nonumber\\
&&\left.-\frac{1}{2}(\partial_{x}\phi_{2})^{2}-\frac{1}{2}(\partial_{t}\phi_{2})^{2}+
\bar{\psi}_{2}\partial_{t}\bar{\psi}_{2}+\psi_{2}\partial_{t}\psi_{2}+V_{2}+W_{2}\right]_{x=0}.\nonumber\\
\end{eqnarray}
Using the boundary conditions (\ref{df1})-(\ref{df4}) and assuming that 
\begin{eqnarray}
\frac{1}{2}(\partial_{\phi_{1}}B_{0})^{2}-\frac{1}{2}(\partial_{\phi_{2}}B_{0})^{2}-V_{1}+V_{2}&=&0,\label{condition1}\\
(\partial_{\phi_{1}}B_{1})^{2}=(\partial_{\phi_{2}}B_{1})^{2}&=&0,\label{condition1,5}
\end{eqnarray}
  equation (\ref{momento1}) takes the following form
\begin{eqnarray}\label{momento2}
\frac{dP}{dt}&=&\Big[-(\partial_{t}\phi_{2}\partial_{\phi_{1}}+\partial_{t}\phi_{1}\partial_{\phi_{2}})(B_{0}+
B_{1})+\partial_{\phi_{1}}B_{0}\partial_{\phi_{1}}B_{1}-\partial_{\phi_{2}}B_{0}\partial_{\phi_{2}}B_{1}\nonumber\\
&&-\partial_{t}\bar{\psi}_{1}\partial_{\bar{\psi}_{1}}B_{1}-\partial_{t}\bar{\psi}_{2}\partial_{\bar{\psi}_{2}}B_{1}+
\partial_{t}\psi_{1}\partial_{\psi_{1}}B_{1}+\partial_{t}\psi_{2}\partial_{\psi_{2}}B_{1}\nonumber\\
&&-W_{1}+W_{2}+\partial_{t}(\bar{\psi}_{1}\bar{\psi}_{2})-\partial_{t}(\psi_{1}\psi_{2})\Big]_{x=0}.
\end{eqnarray}      
Introducing new variables
\begin{eqnarray}
\phi_{\pm}&=&\phi_{1}\pm\phi_{2}\quad\to\quad\left\{\begin{array}{l}\partial_{\phi_{1}}=\partial_{\phi_{+}}+\partial_{\phi_{-}}\\ 
\partial_{\phi_{2}}=\partial_{\phi_{+}}-\partial_{\phi_{-}}\end{array}\right.\\
\bar{\psi}_{\pm}&=&\bar{\psi}_{1}\pm\bar{\psi}_{2}\quad\to\quad\left\{\begin{array}{l}\partial_{\bar{\psi}_{1}}=
\partial_{\bar{\psi}_{+}}+\partial_{\bar{\psi}_{-}}\\
\partial_{\bar{\psi}_{2}}=\partial_{\bar{\psi}_{+}}-\partial_{\bar{\psi}_{-}}\end{array}\right.\\
\psi_{\pm}&=&\psi_{1}\pm\psi_{2}\quad\to\quad\left\{\begin{array}{l}\partial_{\psi_{1}}=\partial_{\psi_{+}}+\partial_{\psi_{-}}\\
\partial_{\psi_{2}}=\partial_{\psi_{+}}-\partial_{\psi_{-}}\end{array}\right.
\end{eqnarray}
we can see from the equations (\ref{df3}) and (\ref{df4}) that
\begin{eqnarray}\label{condition2}
\partial_{\psi_{+}}B_{1}=0, \qquad \partial_{\bar{\psi}_{-}}B_{1}=0.
\end{eqnarray}
The above conditions suggest that $B_{1}$ is independent of $\psi_{+}$ and $\bar{\psi}_{-}$.\\
Let us assume that 
\begin{eqnarray}\label{condition3}
\partial_{\phi_{+}}\partial_{\phi_{-}}B_{0}=0, \qquad \partial_{\phi_{+}}\partial_{\phi_{-}}B_{1}=0, 
\qquad \partial_{\bar{\psi}_{+}}\partial_{\psi_{-}}B_{1}=0,
\end{eqnarray}   
so that we  decompose
\begin{eqnarray}
B_{0}&=&B_{0}^{+}(\phi_{+})+B_{0}^{-}(\phi_{-}),\label{B0}\\
B_{1}&=&B_{1}^{+}(\phi_{+},\bar{\psi}_{+},f_{1})+B_{1}^{-}(\phi_{-},\psi_{-},f_{1}).\label{B1}
\end{eqnarray}
In terms of the new variables and using (\ref{B0}), (\ref{B1}) and (\ref{df5}), 
 equation (\ref{momento2}) is written as
\begin{eqnarray}
\frac{dP}{dt}&=&\Big[\partial_{t}(-B_{0}^{+}+B_{0}^{-}-B_{1}^{+}+B_{1}^{-}+
\bar{\psi}_{1}\bar{\psi}_{2}-\psi_{1}\psi_{2})\Big]_{x=0}\nonumber\\
&+&\Big[\frac{1}{2}\partial_{f_{1}}B_{1}^{+}\partial_{f_{1}}B_{1}^{-}+
2\partial_{\phi_{+}}B_{0}^{+}\partial_{\phi_{-}}B_{1}^{-}+2\partial_{\phi_{-}}B_{0}^{-}\partial_{\phi_{+}}B_{1}^{+}
-W_{1}+W_{2}\Big]_{x=0},
\end{eqnarray}  
which reduces to a total time derivative, provided
\begin{eqnarray}\label{condition4}
\frac{1}{2}\partial_{f_{1}}B_{1}^{+}\partial_{f_{1}}B_{1}^{-}+
2\partial_{\phi_{+}}B_{0}^{+}\partial_{\phi_{-}}B_{1}^{-}+2\partial_{\phi_{-}}B_{0}^{-}\partial_{\phi_{+}}B_{1}^{+}=W_{1}-W_{2}.
\end{eqnarray}
Thus, the combination
\begin{eqnarray}
{\mathcal{P}}=P+\Big[(B_{0}^{+}-B_{0}^{-})+(B_{1}^{+}-B_{1}^{-})-\bar{\psi}_{1}\bar{\psi}_{2}+\psi_{1}\psi_{2})\Big]_{x=0}
\end{eqnarray} 
is    conserved.
For the energy:
\begin{eqnarray}
E&=&\int_{-\infty}^{0}dx\left[\frac{1}{2}(\partial_{x}\phi_{1})^{2}+\frac{1}{2}(\partial_{t}\phi_{1})^{2}-
\bar{\psi}_{1}\partial_{x}\bar{\psi}_{1}+\psi_{1}\partial_{x}\psi_{1}+V_{1}+W_{1}\right]\nonumber\\
&+&\int_{0}^{\infty}dx\left[\frac{1}{2}(\partial_{x}\phi_{2})^{2}+\frac{1}{2}(\partial_{t}\phi_{2})^{2}-
\bar{\psi}_{2}\partial_{x}\bar{\psi}_{2}+\psi_{2}\partial_{x}\psi_{2}+V_{2}+W_{2}\right],\nonumber\\
\end{eqnarray}
the  conserved quantity is given by  the combination
\begin{eqnarray}
{\mathcal E}=E+\Big[(B_{0}^{+}+B_{0}^{-})+(B_{1}^{+}+B_{1}^{-})-\bar{\psi}_{1}\bar{\psi}_{2}-\psi_{1}\psi_{2}\Big]_{x=0}.
\end{eqnarray}

\subsection{Bosonic free field theory}
 Consider the case where the fermionic fields  vanish and let us recall the results of ref. \cite{bowcock2}.
The Lagrangian density in the 
regions $x<0$ and $x>0$ are given by
\begin{eqnarray}
{\mathcal{L}}_{p}&=&\frac{1}{2}(\partial_{x}\phi_{p})^{2}-\frac{1}{2}(\partial_{t}\phi_{p})^{2}+V_{p}, \quad p=1,2      
\end{eqnarray}
where 
\begin{eqnarray}
V_{p}=\frac{1}{2}m^{2}\phi_{p}^{2}, \qquad p=1,2.
\end{eqnarray}
The field equations are
\begin{eqnarray}
\partial_{x}^{2}\phi_{p}-\partial_{t}^{2}\phi_{p}=m^{2}\phi_{p} \qquad p=1,2.
\end{eqnarray}
At $x=0$ the Lagrangian density associated with the defect is 
\begin{eqnarray}
{\mathcal{L}}_{D}&=&\frac{1}{2}\left(\phi_{2}\partial_{t}\phi_{1}-\phi_{1}\partial_{t}\phi_{2}\right)+B_{0}(\phi_{1},\phi_{2}).
\end{eqnarray} 
A solution satisfying the equation (\ref{condition1}) and the first equation of (\ref{condition3}) is given by
\begin{eqnarray}
B_{0}=-\frac{m\beta^{2}}{4}(\phi_{1}-\phi_{2})^{2}-\frac{m}{4\beta^{2}}(\phi_{1}+\phi_{2})^{2},
\end{eqnarray}   
where $\beta$ is a free parameter.
Thus, the defect conditions at $x=0$, namely (\ref{df1}) and (\ref{df2}) become 
\begin{eqnarray}
\partial_{x}\phi_{1}-\partial_{t}\phi_{2}&=&\frac{m\beta^{2}}{2}(\phi_{1}-\phi_{2})+\frac{m}{2\beta^{2}}(\phi_{1}+\phi_{2}),\nonumber\\
\partial_{x}\phi_{2}-\partial_{t}\phi_{1}&=&\frac{m\beta^{2}}{2}(\phi_{1}-\phi_{2})-\frac{m}{2\beta^{2}}(\phi_{1}+\phi_{2}).
\end{eqnarray} 
The modified conserved momentum and energy are, respectively
\begin{eqnarray}
\mathcal{P}&=&P+\left[-\frac{m\beta^{2}}{4}(\phi_{1}-\phi_{2})^{2}+\frac{m}{4\beta^{2}}(\phi_{1}+\phi_{2})^{2},\right]_{x=0},\nonumber\\
\mathcal{E}&=&E+\left[-\frac{m\beta^{2}}{4}(\phi_{1}-\phi_{2})^{2}-\frac{m}{4\beta^{2}}(\phi_{1}+\phi_{2})^{2},\right]_{x=0}
\end{eqnarray}

\subsection{Fermionic free field theory}
Let us consider the following Lagrangian
\begin{eqnarray}
{\mathcal{L}}_{p}=-\bar{\psi}_{p}\partial_{x}\bar{\psi}_{p}+\bar{\psi}_{p}\partial_{t}\bar{\psi}_{p}+
\psi_{p}\partial_{x}\psi_{p}+\psi_{p}\partial_{t}\psi_{p}+W_{p}(\psi_{p},\bar{\psi}_{p}),
\end{eqnarray}
where,
\begin{eqnarray}
W_{p}=2m\bar{\psi}_{p}\psi_{p}, \quad p=1,2.
\end{eqnarray} 
Then, the field equations in the regions $x<0$ and $x>0$ are
\begin{eqnarray}
\partial_{x}\psi_{p}+\partial_{t}\psi_{p}=m\bar{\psi}_{p}, \qquad \partial_{x}\bar{\psi}_{p}-
\partial_{t}\bar{\psi}_{p}=m\psi_{p}, \qquad p=1,2
\end{eqnarray}
The Lagrangian associated with the defect is taken to be
\begin{eqnarray}
{\mathcal{L}}_{D}&=&-\psi_{1}\psi_{2}-\bar{\psi}_{1}\bar{\psi}_{2}+2f_{1}\partial_{t}f_{1}+
B_{1}(\psi_{1},\psi_{2},\bar{\psi}_{1},\bar{\psi}_{2},f_{1}).
\end{eqnarray}
where
\begin{eqnarray}
B_{1}=-\frac{2i}{\beta}\sqrt{m}f_{1}(\bar{\psi}_{1}+\bar{\psi}_{2})+i\beta\sqrt{m}f_{1}(\psi_{1}-\psi_{2}),
\end{eqnarray}
satisfies the condition required by the conservation of the modified momentum and 
energy (\ref{condition3}) and  (\ref{condition4}), i.e.,
\begin{eqnarray}
\partial_{\bar{\psi}_{+}}\partial_{\psi_{-}}B_{1}=0,\quad \quad 
\frac{1}{2}\partial_{f_{1}}B_{1}^{+}\partial_{f_{1}}B_{1}^{-}=W_{1}-W_{2}.
\end{eqnarray}
The defect conditions at $x=0$ are then 
\begin{eqnarray}
\psi_{1}+\psi_{2}=-i\beta\sqrt{m}f_{1},\quad \quad 
\bar{\psi}_{1}-\bar{\psi}_{2}=-\frac{2i}{\beta}\sqrt{m}f_{1},\nonumber\\
\partial_{t}f_{1}=\frac{i}{2\beta}\sqrt{m}(\bar{\psi}_{1}+\bar{\psi}_{2})-\frac{i\beta}{4}\sqrt{m}(\psi_{1}-\psi_{2}).
\end{eqnarray}
and the modified conserved  momentum and  energy are  given by
\begin{eqnarray}
\mathcal{P}&=&P+\Big[-\frac{2i}{\beta}\sqrt{m}f_{1}(\bar{\psi}_{1}+\bar{\psi}_{2})-
i\beta\sqrt{m}f_{1}(\psi_{1}-\psi_{2})-\bar{\psi}_{1}\bar{\psi}_{2}+\psi_{1}\psi_{2}\Big]_{x=0}\nonumber\\
\mathcal{E}&=&E+\Big[-\frac{2i}{\beta}\sqrt{m}f_{1}(\bar{\psi}_{1}+\bar{\psi}_{2})+
i\beta\sqrt{m}f_{1}(\psi_{1}-\psi_{2})-\bar{\psi}_{1}\bar{\psi}_{2}-\psi_{1}\psi_{2}\Big]_{x=0}.\nonumber\\
\end{eqnarray}

\subsection{Supersymmetric Sinh-Gordon}
Consider the Lagragian density (\ref{lll}) with 
\begin{eqnarray}
{\mathcal{L}}_{p}&=&\frac{1}{2}(\partial_{x}\phi_{p})^{2}-\frac{1}{2}(\partial_{t}\phi_{p})^{2}-\bar{\psi}_{p}\partial_{x}\bar{\psi}_{p}+\bar{\psi}_{p}\partial_{t}\bar{\psi}_{p}+\psi_{p}\partial_{x}\psi_{p}+\psi_{p}\partial_{t}\psi_{p}\nonumber\\
&&+V_{p}(\phi_{p})+W_{p}(\phi_{p},\psi_{p},\bar{\psi}_{p}), \qquad p=1,2\nonumber\\
{\mathcal{L}}_{D}&=&\frac{1}{2}\left(\phi_{2}\partial_{t}\phi_{1}-\phi_{1}\partial_{t}\phi_{2}\right)-\psi_{1}\psi_{2}-\bar{\psi}_{1}\bar{\psi}_{2}+2f_{1}\partial_{t}f_{1}\nonumber\\
&&B_{0}(\phi_{1},\phi_{2})+B_{1}(\phi_{1},\phi_{2},\psi_{1},\psi_{2},\bar{\psi}_{1},\bar{\psi}_{2},f_{1}),
\end{eqnarray}
where,
\begin{eqnarray}
V_{p}=4m^{2}\cosh(2\phi_{p}), \qquad W_{p}=8m\,\bar{\psi}_{p}\psi_{p}\cosh\phi_{p} \qquad p=1,2
\end{eqnarray}
and
\begin{eqnarray}
B_{0}&=&-m\beta^{2}\cosh(\phi_{1}-\phi_{2})-\frac{4m}{\beta^{2}}\cosh(\phi_{1}+\phi_{2}),\label{bezero}\\
B_{1}&=&-\frac{4i}{\beta}\sqrt{m}\,\cosh\left(\frac{\phi_{1}+\phi_{2}}{2}\right)f_{1}(\bar{\psi}_{1}+\bar{\psi}_{2})\nonumber\\
&&+2i\beta\sqrt{m}\,\cosh\left(\frac{\phi_{1}-\phi_{2}}{2}\right)f_{1}(\psi_{1}-\psi_{2}).\label{beum} 
\end{eqnarray}
The field equations are

$x<0$:
\begin{eqnarray}\label{mov1}
\partial_{x}^{2}\phi_{1}-\partial_{t}^{2}\phi_{1}&=&8m\sinh(2\phi_{1})+8m\,\bar{\psi}_{1}\psi_{1}\sinh\phi_{1},\nonumber\\
(\partial_{x}-\partial_{t})\bar{\psi}_{1}&=&4m\,\psi_{1}\cosh\phi_{1},\nonumber\\
(\partial_{x}+\partial_{t})\psi_{1}&=&4m\,\bar{\psi}_{1}\cosh\phi_{1},
\end{eqnarray}
$x>0$:
\begin{eqnarray}\label{mov2}
\partial_{x}^{2}\phi_{2}-\partial_{t}^{2}\phi_{2}&=&8m\sinh(2\phi_{2})+8m\,\bar{\psi}_{2}\psi_{2}\sinh\phi_{2},\nonumber\\
(\partial_{x}-\partial_{t})\bar{\psi}_{2}&=&4m\,\psi_{2}\cosh\phi_{2},\nonumber\\
(\partial_{x}+\partial_{t})\psi_{2}&=&4m\,\bar{\psi}_{2}\cosh\phi_{2},
\end{eqnarray}
$x=0$:
\begin{eqnarray}
\partial_{x}\phi_{1}-\partial_{t}\phi_{2}&=&m\beta^{2}\sinh(\phi_{1}-\phi_{2})+\frac{4m}{\beta^{2}}\sinh(\phi_{1}+\phi_{2})\nonumber\\
&&\frac{2i}{\beta}\sqrt{m}\,\sinh\left(\frac{\phi_{1}+\phi_{2}}{2}\right)f_{1}(\bar{\psi}_{1}+\bar{\psi}_{2})\nonumber\\
&&-i\beta\sqrt{m}\,\sinh\left(\frac{\phi_{1}-\phi_{2}}{2}\right)f_{1}(\psi_{1}-\psi_{2}).
\label{defeito1}\\
\partial_{x}\phi_{2}-\partial_{t}\phi_{1}&=&m\beta^{2}\sinh(\phi_{1}-\phi_{2})-\frac{4m}{\beta^{2}}\sinh(\phi_{1}+\phi_{2})\nonumber\\
&&-\frac{2i}{\beta}\sqrt{m}\,\sinh\left(\frac{\phi_{1}+\phi_{2}}{2}\right)f_{1}(\bar{\psi}_{1}+\bar{\psi}_{2})\nonumber\\
&&-i\beta\sqrt{m}\,\sinh\left(\frac{\phi_{1}-\phi_{2}}{2}\right)f_{1}(\psi_{1}-\psi_{2}).
\label{defeito2}\\ 
\psi_{1}+\psi_{2}&=&-2i\beta\sqrt{m}\,\cosh\left(\frac{\phi_{1}-\phi_{2}}{2}\right)f_{1},
\label{defeito3}\\
\bar{\psi}_{1}-\bar{\psi}_{2}&=&-\frac{4i}{\beta}\sqrt{m}\,\cosh\left(\frac{\phi_{1}+\phi_{2}}{2}\right)f_{1},
\label{defeito4}\\
\partial_{t}f_{1}&=&\frac{i}{\beta}\sqrt{m}\,\cosh\left(\frac{\phi_{1}+\phi_{2}}{2}\right)(\bar{\psi}_{1}+\bar{\psi}_{2})\nonumber\\
&&-\frac{i\beta}{2}\sqrt{m}\,\cosh\left(\frac{\phi_{1}-\phi_{2}}{2}\right)(\psi_{1}-\psi_{2}).
\label{defeito5} 
\end{eqnarray}
Due to (\ref{bezero}) and  (\ref{beum}),  the first order equations (\ref{defeito1})- (\ref{defeito5}) 
agrees with the Backlund transformations of the Appendix.
Applying the operator $(\partial_{x}+\partial_{t})$ to  (\ref{defeito3}) and using equations (\ref{mov1}) and (\ref{mov2}), we obtain
\begin{eqnarray}
\partial_{x}f_{1}&=&\frac{i}{\beta}\sqrt{m}\,\cosh\left(\frac{\phi_{1}+\phi_{2}}{2}\right)(\bar{\psi}_{1}+\bar{\psi}_{2})\nonumber\\
&&+\frac{i\beta}{2}\sqrt{m}\,\cosh\left(\frac{\phi_{1}-\phi_{2}}{2}\right)(\psi_{1}-\psi_{2}).
\label{defeito6} 
\end{eqnarray}
Equations (\ref{bezero}) and (\ref{beum}) satisfy the conditions (\ref{condition1}), (\ref{condition1,5}), 
(\ref{condition2}), (\ref{condition3}) and(\ref{condition4}).\\
The modified conserved momentum and  energy are given by:
\begin{eqnarray}
\mathcal{P}&=&P+\Bigg[-m\beta^{2}\cosh(\phi_{1}-\phi_{2})+\frac{4m}{\beta^{2}}\cosh(\phi_{1}+\phi_{2})\nonumber\\
&&-\frac{4i}{\beta}\sqrt{m}\,\cosh\left(\frac{\phi_{1}+\phi_{2}}{2}\right)f_{1}(\bar{\psi}_{1}+\bar{\psi}_{2})\nonumber\\
&&-2i\beta\sqrt{m}\,\cosh\left(\frac{\phi_{1}-\phi_{2}}{2}\right)f_{1}(\psi_{1}-\psi_{2})-
\bar{\psi}_{1}\bar{\psi}_{2}+\psi_{1}\psi_{2}\Bigg]_{x=0}\nonumber\\
\\
\mathcal{E}&=&E+\Bigg[-m\beta^{2}\cosh(\phi_{1}-\phi_{2})-\frac{4m}{\beta^{2}}\cosh(\phi_{1}+\phi_{2})\nonumber\\
&&-\frac{4i}{\beta}\sqrt{m}\,\cosh\left(\frac{\phi_{1}+\phi_{2}}{2}\right)f_{1}(\bar{\psi}_{1}+\bar{\psi}_{2})\nonumber\\
&&+2i\beta\sqrt{m}\,\cosh\left(\frac{\phi_{1}-\phi_{2}}{2}\right)f_{1}(\psi_{1}-\psi_{2})-
\bar{\psi}_{1}\bar{\psi}_{2}-\psi_{1}\psi_{2}\Bigg]_{x=0}
\end{eqnarray} 
For simplicity, we shall take $m=1$ from now on.

\section{Backlund Solutions for the super Sinh-Gordon}
In this section we discuss the various solutions compatible with the Backlund 
transformation (\ref{defeito1}) -  (\ref{defeito5}) at $x=0$.

\begin{itemize}
\item {\bf {vacuum- 1-boson system}}
Let   $\phi_1  = 0$, $\bar \psi_1 = \bar \psi_2 =  0$ and $\phi_2 \neq 0$.  From (\ref{defeito1})
and  (\ref{defeito2}) we find
\br
\phi_2 = ln \( {{1+{1\o 2} b_1 \rho(\s)}\o {1-{1\o 2} b_1 \rho(\s)}}\), \quad \quad \rho(\s) = \exp (2\s(x+t) +{2\o \s}(x-t))
\label{g.2}
\er
where $\s =-{{2}\o {\b^2}}$ and $b_1$ is an arbitrary constant.

\item {\bf {vacuum- 1-fermion system}}
We consider solutions were $\phi_1  = \phi_2  = 0$ and $ \psi_1 =\bar \psi_1=0$ and $\bar \psi_2\neq 0$.  
From (\ref{defeito3}) and  (\ref{defeito4}) we find
\br
\psi_2 = -2i\b f_1, \quad \quad \bar \psi_2 = {{4i}\o {\b}}f_1.
\label{f1}
\er
Henceforth $\psi_2 = {{1}\o {\s}} \bar \psi_2$.  Eliminating $f_1$ and $\psi_2$ in (\ref{defeito5}) and  (\ref{defeito6}) we arrive at
\br
\pa_x \bar \psi_2 = 2(\s +{{1}\o {\s}}) \bar \psi_2, \quad \quad \pa_t \bar \psi_2 = 2(\s -{{1}\o {\s}}) \bar \psi_2
\label{psi2}
\er
The solution is then given by
$\bar \psi_2 = {{ 2c_1 \s }\o {\rho(\s)}}$.
where $\rho(\s)$ is given in (\ref{g.2}) and $c_1$ is an arbitrary grassmanian constant.

\item {\bf{vacuum- fermion/boson system}}
We consider solutions were $\phi_1  = \psi_1 = \bar \psi_1 = 0$. 
 From (\ref{defeito1}) and  (\ref{defeito2}) we find
 \br
 \pa_t \phi_2 &=& -{{2i}\o {\b}}f_1 \sinh ({{\phi_2}\o 2}) \bar \psi_2 + 
 i\b f_1 \sinh ({{\phi_2}\o 2})  \psi_2 -2 ({{2}\o {\b^2}} - {{\b^2}\o 2})\sinh ({{\phi_2}}) ,\nonu \\
  \pa_x \phi_2 &=& -{{2i}\o {\b}}f_1 \sinh ({{\phi_2}\o 2}) \bar \psi_2 - 
 i\b f_1 \sinh ({{\phi_2}\o 2})  \psi_2 -2({{2}\o {\b^2}} +{{\b^2}\o 2})\sinh ({{\phi_2}})
 \label{yy}
 \er
 Relations (\ref{defeito3}) and  (\ref{defeito4})  yields 
 \br
 \psi_2 = -2i\b \cosh ({{\phi_2}\o 2}) f_1, \quad \bar \psi_2 = {{4i}\o {\b}} \cosh ({{\phi_2}\o 2}) f_1
 \label{ff1}
 \er
 and henceforth 
 $\psi_2 = {{1}\o {\s}} \bar \psi_2$.  Eliminating $f_1$ from (\ref{ff1}) into    (\ref{yy})we find
 \br
 \pa_t \phi_2 = 2 (\s -{{1}\o {\s}})\sinh ({{\phi_2}}), \quad \pa_x \phi_2 = 2 (\s +{{1}\o {\s}})\sinh ({{\phi_2}})
 \label{zz}
 \er
 with solution given by
 \br
 \phi_2 = ln \( {{1+{1\o 2} b_1 \rho(\s)}\o {1-{1\o 2} b_1 \rho(\s)}}\).
 \label{pphi}
 \er
 Substituting now $f_1$ in terms of $\bar \psi_2, \psi_2$  and $\phi_2$ from (\ref{pphi}), we find as solution of  
 (\ref{defeito5}) and  (\ref{defeito6}) 
 \br
 \bar \psi_2 = -2c_1 \g \rho \( {{1}\o {1-{1\o 4} b_1^2 \rho(\s)^2}}\)
 \label{g.5}
 \er
and similar for $\psi_2$.  Similar results are obtained interchanging $\phi_1, \psi_1, \bar \psi_1 
\rightarrow \phi_2, \psi_2, \bar \psi_2$.

Let us now change variables $\phi_p \rightarrow {{i}\o {2}}\phi_p, 
\;\; x \rightarrow x/4, \;\;  t \rightarrow t/4$  in order to discuss the solutions of super sine-Gordon model.

\item {\bf {boson - boson system}}
Let us now consider solutions with $\bar \psi_1 = \bar \psi_2 = 0, \; \phi_1 \neq 0, \; \phi_2 \neq 0$. 
This case yields precisely the
solution obtained in \cite{bowcock1},\cite{bowcock2}.
\br
e^{i\phi_a/2} = {{1-iE_a}\o {1+iE_a}}, \quad \quad E_a = R_ae^{\a_a x+ \b_a t}, \quad 
\label{g.8} 
\er
where $\a_a^2 -\b_a^2 =1, \;\; a=1,2$.
The Backlund transformation (\ref{defeito1}) and (\ref{defeito2}) 
\br
\pa_x \phi_1 - \pa_t \phi_2 = -\s \sin ({{\phi_1+\phi_2}\o {2}}) - {{1}\o {\s}}\sin ({{\phi_1-\phi_2}\o {2}}), \nonu \\
\pa_x \phi_2 - \pa_t \phi_1 = \s \sin ({{\phi_1+\phi_2}\o {2}}) - {{1}\o {\s}}\sin ({{\phi_1-\phi_2}\o {2}})
\label{g.7}
\er
imply the following relation for $\a_a = \cosh \theta_a$ and $\b_a = \sinh \theta_a$
\br
\theta_1 = \theta_2 = \theta, \quad \quad R_2 = \({{e^{\theta} +\s}\o {e^{\theta} -\s}}\)R_1.
\label{g.8a}
\er
The defect preserves the soliton velocity, allowing at most a  phase shift. 
However, when $\s <0$ and for $e^{\theta }= |\s |$ we find $\phi_2 =0$. 
This configuration corresponds to an absortion of the soliton.
On the other hand, if $\s >0$ and  $e^{\theta }= |\s |$, $\phi_1 =0$ corresponding to an 
emission of the soliton by the defect.

\item {\bf {Fermion - Fermion system}}
We consider solutions were $\phi_1 = \phi_2 =  0$.  The solution of eqns. (\ref{mov1}) and (\ref{mov2})  are of the form 
\br
\bar \psi_a = \eps S_a \exp (\a_a x + \b_a t),  \quad    \psi_a = e^{-\theta_a} \bar \psi_a,    \quad a=1,2
\label{g.9a}
\er
where $\eps$ is a grassmanian parameter and $\a_a^2 - \b_a^2 =1$.  From  equation (\ref{defeito3})-(\ref{defeito6}) we find
\br
{{2}\o {\b}} \(S_2 \b_1 e^{-\theta_1} e^{\a_1x+\b_1t} +S_2 \b_2 e^{-\theta_2}e^{\a_2x+\b_2t}\) &=&
{{1}\o {\b}} \(S_1 e^{\a_1x+\b_1t} +S_2 e^{\a_2x+\b_2t}\) \nonu \\
&-& {{\b}\o {2}}\(S_1  e^{-\theta_1} e^{\a_1x+\b_1t} -S_2  e^{-\theta_2}e^{\a_2x+\b_2t}\) \nonu \\
\label{xxx}
\er
For $\a_1 \neq \a_2$, it follows that $e^{-\theta_1} = - e^{-\theta_2} = {{\b^2}\o 2}$.  For $\a_1 = \a_2$, we find $\theta_1 = \theta_2
=\theta$ and 
\br
 S_2 = \({{e^{\theta} + \s }\o {e^{\theta} - \s}}\)S_1
\label{g.11}
\er
where $\a_a = \cosh \theta $, $\b_a = \sinh \theta, \;\; $ 
Again the velocity is preserved with only a phase shift being  allowed when the soliton interacts with the defect.
Notice that limiting cases, where $S_2=0$ or $S_1=0$  are obtained when  $\s <0$ and $e^{\theta} = |\s |$ or
 $\s >0$ and $e^{\theta} = \s $.  These cases correspond to total absortion or creation of one fermion respectively.
 
\item {\bf {Fermion/Boson - Fermion/Boson system}}

Consider  the following solution of eqns. (\ref{mov1}) and (\ref{mov2})
\br
e^{i\phi_a} &=& {{1-iE_a}\o{{1+iE_a}}}, \quad \quad E_a = R_a e^{\a_ax +\b_a t}, \quad \a_a^2-\b_a^2 = 1, a=1,2
\nonu \\
\bar \psi_a &=& \eps S_a e^{\a_ax +\b_a t}\({{1}\o {1+iE_a}} + {{1}\o {1-iE_a}}\), 
\quad \quad   \psi_a = e^{-\theta_a} \bar \psi_a \nonu \\
 \label{g.11a}
\er
We now substitute (\ref{g.9a}) and (\ref{g.11}) into  equations (\ref{defeito1}) - (\ref{defeito6}), 
\br
\pa_x \phi_1 -\pa_t \phi_2 &=&-\s \sin ({{\phi_1+\phi_2}\o {2}}) - {{1}\o {\s}} \sin ({{\phi_1-\phi_2}\o {2}}), \nonu \\
\pa_x \phi_2 -\pa_t \phi_1 &=&\s \sin ({{\phi_1+\phi_2}\o {2}}) - {{1}\o {\s}} \sin ({{\phi_1-\phi_2}\o {2}}), \nonu \\
({\psi_1+\psi_2}) \cos ({{\phi_1+\phi_2}\o {4}})
&=& -{{1}\o {{\s}}} ({\bar \psi_1-\bar \psi_2})\cos ({{\phi_1-\phi_2}\o {4}}). 
\label{g.9}
\er
Writing  $\a_a = \cosh \theta_a, \quad \b_a = \sinh \theta_a $ in (\ref{g.11a}) we find from (\ref{g.9})
\br
\theta_1 = \theta_2 = \theta, \quad \quad S_2 = \({{e^{\theta} + \s }\o {e^{\theta} - \s}}\)S_1, \quad \quad 
R_2 = \({{e^{\theta} + \s }\o {e^{\theta} - \s}}\)R_1
\er
Again the velocity is preserved with only a phase shift being  allowed when the soliton interacts with the defect.
Notice that limiting cases, where $S_2 =R_2=0$ or $S_1=R_1=0$  are obtained when  $\s <0$ and $e^{\theta} = |\s |$ or
 $\s >0$ and $e^{\theta} = \s $.  These cases correspond to total absortion or creation of one soliton respectively.
 
\end{itemize}

\section{Zero Curvature  Formulation}

Consider the $sl(2,1)$ super Lie algebra with generators 
\br
h_1= \a_1 \cdot H, \;\; h_2 = \a_2 \cdot H,\;\;  E_{\pm \a_1}, \;\; E_{\pm \a_2},\;\; E_{\pm (\a_1+\a_2)}
\label{0.1}
\er
where $\a_1$ and $\a_2, \a_1 + \a_2$ are bosonic and fermionic roots respectively. We now extend the finite 
dimensional Lie algebra $sl(2,1)$ to an 
affine structure by introducing the spectral parameter $\l$ as follows,
\br
T_a \rightarrow T_a^{(n)} = \l^n T_a, \quad \l \in C
\label{xx}
\er  
where $n\in Z$  or $n \in Z+ 1/2$ according to $T_a$  denoting bosonic or fermionic generators respectively \cite{npb}.

The super sinh-Gordon model can be  described by the Lax pair (see for instance ref. \cite{npb})
\br
a_t^{(p)} &=& -{1\o 2} \pa_x \phi_{p} h_1 + (\l - {1\o \l}) (h_1 +2h_2) + ({e^{\phi_{p}}\o \l} - \l e^{-\phi_{p}})E_{\a_1} + 
({e^{-\phi_{p}}\o \l} - \l e^{\phi_{p}})E_{-\a_1} \nonu \\
&+& 
(\l^{{1\o 2}}e^{{1\o 2}\phi_{p}}\bar \psi_p  +  \l^{-{1\o 2}}e^{-{1\o 2}\phi_{p}}\psi_p )E_{\a_2 } + 
(\l^{{1\o 2}}e^{-{1\o 2}\phi_{p}}\bar \psi_p  -  \l^{-{1\o 2}}e^{{1\o 2}\phi_{p}}\psi_p )E_{-\a_2 }\nonu \\
&+& 
(\l^{{1\o 2}}e^{-{1\o 2}\phi_{p}}\bar \psi_{p}  +  \l^{-{1\o 2}}e^{{1\o 2}\phi_{p}}\psi_{p} )E_{\a_1+\a_2 } + 
(\l^{{1\o 2}}e^{{1\o 2}\phi_{p}}\bar \psi_{p}  -  \l^{-{1\o 2}}e^{-{1\o 2}\phi_{p}}\psi_{p} )E_{-\a_1-\a_2 }\nonu 
\label{2.1}
\er
\br
a_x^{(p)} &=& -{1\o 2} \pa_t \phi_{p} h_1 + (\l + {1\o \l}) (h_1 +2h_2) - ({e^{\phi_{p}}\o \l} + \l e^{-\phi_{p}})E_{\a_1} - 
({e^{-\phi_{p}}\o \l} + \l e^{\phi_{p}})E_{-\a_1} \nonu \\
&+& 
(\l^{{1\o 2}}e^{{1\o 2}\phi_{p}}\bar \psi_{p}  -  \l^{-{1\o 2}}e^{-{1\o 2}\phi_{p}}\psi_p )E_{\a_2 } + 
(\l^{{1\o 2}}e^{-{1\o 2}\phi_{p}}\bar \psi_{p}  +  \l^{-{1\o 2}}e^{{1\o 2}\phi_{p}}\psi_{p} )E_{-\a_2 }\nonu \\
&+& 
(\l^{{1\o 2}}e^{-{1\o 2}\phi_{p}}\bar \psi_{p}  -  \l^{-{1\o 2}}e^{{1\o 2}\phi_{p}}\psi_{p} )E_{\a_1+\a_2 } + 
(\l^{{1\o 2}}e^{{1\o 2}\phi_{p}}\bar \psi_{p}  +  \l^{-{1\o 2}}e^{-{1\o 2}\phi_{p}}\psi_{p} )E_{-\a_1-\a_2 } \nonu \\
p=1,2
\label{2.2}
\er
where the power of $\l$ denotes the effective grading  and 
henceforth the generators $h_i, E_{\pm \a}$ correspond to $h_i^{(0)}, E_{\pm \a}^{(0)}$ respectively. 
 The corresponding zero curvature
equation  yields the equations of motion (\ref{mov1}) and (\ref{mov2}) with $m=1$.

In order to describe the integrability of the system, we follow ref. \cite{corrigan}  and split the space into 
two overlaping regions, namely, $x\leq b$ and $x\geq a$ with $a<b$.  Inside the overlap region, i.e.,  
$ a\leq x \leq b$,  define the Lax pair to be,
\br
\hat {a}_t^{(1)} &=& a_t^{(1)} -{1\o 2} \theta(x-a) ( (\pa_x \phi_1 -\pa_t \phi_2  +\pa_{\phi_1}B_0 + \pa_{\phi_1}B_1)h_1 \nonu \\
&+& (\psi_1 + \psi_2 -\pa_{\psi_1} B_1)E_1 + (\pa_t f_1 + {1\o 4} \pa_{f_1}B_1)E_1^{\pr}), \nonu \\
\hat {a}_x^{(1)} &=& \theta(a-x) a_x^{(1)}, \nonu \\
\hat {a}_t^{(2)} &=& a_t^{(2)} -{1\o 2} \theta(b-x) ( (\pa_x \phi_2 -\pa_t \phi_1 -\pa_{\phi_2}B_0 - \pa_{\phi_2}B_1)h_1  \nonu \\
&+& (\bar \psi_1 - \bar \psi_2 + \pa_{\bar \psi_2} B_1)E_2 + (\pa_t f_1 + {1\o 4} \pa_{f_1}B_1)E_2^{\pr}), \nonu \\
\hat {a}_x^{(2)} &=& \theta(x-b) a_x^{(2)}. 
\label{2.3}
\er
where $E_i$ and $ E_i^{\pr}$ denote a pair of  independent  fermionic step operators of $sl(2,1)$.
Within the overlap region, the Lax pairs denoted by the suffices $p=1,2$ are related by gauge transformation. 
In particular for the time component $a_t^{(p)}$, 
\br
K^{-1} \pa_t K = \hat {a}_t^{(2)} - K^{-1} \hat {a}_t^{(1)} K
\label{2.4}
\er
If we now decompose $K$ into 
\br
K =  e^{{1\o 2}\phi_2 h_1} \bar K e^{-{1\o 2}\phi_1 h_1}, \quad \quad \bar K = e^{-m(-1/2) - m(-1)- \cdots }
\label{2.5}
\er
 we find
\br
m(-1/2) &=& {{2if_1}\o {\b}} {\l}^{-{1\o 2}}\(E_{\a_2} -E_{-\a_2} + E_{\a_1+ \a_2} -E_{-\a_1-\a_2}\)  \nonu \\
m(-1) &=& c{\l}^{-1} ( h_1+2h_2 ) - {{2{\l}^{-1}}\o {\b^2}} (E_{\a_1} +E_{-\a_1})
\label{2.6}
\er
provided  the Backlund transformation (\ref{mov1})- (\ref{defeito6}) holds and $c$ is an 
arbitrary constant which, for convenience is chosen as 
$c = {{2}\o {\b^2}}$.  It then follows that $   m(-3/2) = m(-2) = 0$, 
\br
m(-5/2) = {{1}\o {3} } ({{2}\o {\b^2}} )^2 \l ^{-2} m(-1/2), \quad \quad  
m(-3) = {{1}\o {3} } ({{2}\o {\b^2}} )^2 \l ^{-2} m(-1), \cdots 
\label{2.7}
\er
A general closed solution can be put in the following form,
\br
\bar K = I (1+ {{2}\o {\l \b^2}}) \L - m(-1/2)\L - m(-1)\L
\label{2.8}
\er
where $\L $ is an arbitrary constant.  By choosing $\L = (1+ {{2}\o {\l \b^2}})^{-1} $ we recover  
(\ref{2.6})  and (\ref{2.7}) 
after expansion in powers of ${{1} \o {\l}}$ of $\L$,
\br
\bar K = I - m(-1/2)  + {{2}\o {\l \b^2}}m(-1/2) + \cdots  - m(-1) + {{2}\o {\l \b^2}}m(-1) + \cdots 
\label{2.9}
\er
which can be rewritten as in (\ref{2.5}), i.e., 
\br
\bar K = \exp {\( - m(-1/2)- m(-1) -{{1}\o {3} } ({{2}\o {\b^2}} )^2 \l ^{-2} m(-1/2) - \cdots \)}
\label{2.10}
\er
provided  the following relations hold
\br
{{2}\o {\l \b^2}}m(-1)&=& {1\o 2} m(-1)m(-1)\nonu \\
{{2}\o {\l \b^2}}m(-1/2) &=& {1\o 2} \( m(-1/2) m(-1) +  m(-1) m(-1/2)\), \nonu \\
({{2}\o {\l \b^2}})^2 m(-1/2) &=& {1\o 4} (m(-1) m(-1/2) m(-1) + m(-1)m(-1) m(-1/2) \nonu \\ 
&+& m(-1/2) m(-1)m(-1))
\label{2.11}
\er
These are verified using $3$ dimensional matrix representation  of the algebra $sl(2,1)$.

The existence of the gauge transformation (\ref{2.4})  provides  a generating function for an infinite set of
constants of motion (see \cite{bowcock2}) strongly indicating the integrability of the system.

\section*{\sf Acknowledgments}
We are grateful to H. Aratyn for discussions.
LHY acknowledges support from Fapesp,
JFG and AHZ thank CNPq for a partial support.


\section{Appendix  - Backlund Transformation in components}

Let $\Phi$ be a  bosonic super field
\br
\Phi = \phi + \theta_1 \bar \psi + i \theta_2 \psi - \theta_1 \theta_2 F
\label{e.1}
\er
and 
\br
D_z = \pa_{\theta_1} + \theta_1 \pa_z,  \quad D_{\bar z} = \pa_{\theta_2} + \theta_2 \pa_{\bar z}, \quad 
D_z^2 = \pa_z, \quad D^2_{\bar z} = \pa_{\bar z}, \quad D_z D_{\bar z} =-  D_{\bar z}D_z
\label{e.2}
\er
are the corresponding supersymmetric covariant derivatives.  
Here the space-time is defined in terms of the light cone coordinates as 
 $x = {1\o 2}(z+\bar z), \;\; t = {1\o 2}(z-\bar z)$ and therefore 
 $\pa_z = {1\o 2}(\pa_x+\pa_t), \;\; \pa_{\bar z} = {1\o 2}(\pa_x-\pa_t)$.
The equation of motion in 
terms of the superfield (\ref{e.1}) is 
\br
D_z D_{\bar z} \Phi = 2i \sinh \Phi
\label{e.3}
\er
  if the auxiliary field $F$ satisfies $ F = 2i \sinh \phi$. 
  According to Chaichian and Kulish \cite{chai},  the Backlund tranformation for the supersymmetric sinh-Gordon is given by 
   the following first order equations 
 \br
 D_z \Phi_1 &=& D_z \Phi_2 -{{4i}\o {\b}} f \cosh ({{{\Phi_1+\Phi_2}\o 2}})\nonu \\
 D_{\bar z} \Phi_1 &=& -D_{\bar z} \Phi_2 +2\b  f \cosh ({{{\Phi_1-\Phi_2}\o 2}})
 \label{e.4}
\er
Their compatibility 
 imply in (\ref{e.3}) if the fermionic super field $f$ satisfies
 \br
 D_{\bar z} f = \b \sinh ({{{\Phi_1-\Phi_2}\o 2}}), \quad \quad  
D_{ z}f = {{2i}\o {\b}}\sinh ({{{\Phi_1+\Phi_2}\o 2}})
\label{e.5}
\er
Let $f$ be written as $f = f_1 + \theta_1 b_1 +  \theta_2 b_2 + \theta_1 \theta_2 f_2$ where $b_{1}$, $b_{2}$ 
are bosonic fields and $f_{1}$, $f_{2}$ are fermionic fields and denote 
\begin{eqnarray}
\phi_{\pm}&=&\phi_1 \pm {\phi_2}, \quad 
F_{\pm}=F_1 \pm {F_2},\quad \bar{\psi}_{\pm}=\bar{\psi_1} \pm {\bar{\psi_2}}, 
\quad \psi_{\pm}=\psi_1 \pm {\psi_2}.
\end{eqnarray}
 In components, (\ref{e.4}) and (\ref{e.5}) can be written as
\br
f_{1}&=&-\frac{\beta\bar{\psi}_{-}}{4i\cosh\left(\frac{\phi_{+}}{2}\right)}=\frac{i\psi_{+}}{2\beta\cosh\left(\frac{\phi_{-}}{2}\right)},\\
\partial_{\bar{z}}f_{1}&=&\frac{i\beta}{2}\,\mathrm{cosh}\left(\frac{\phi_{-}}{2}\right)\,\psi_{-},\\
\partial_{z}f_{1}&=&\frac{i}{\beta}\,\mathrm{cosh}\left(\frac{\phi_{+}}{2}\right)\,\bar{\psi}_{+},\\
b_{1}&=&\frac{2i}{\beta}\,\mathrm{sinh}\left(\frac{\phi_{+}}{2}\right),\\
\partial_{\bar{z}}b_{1}&=&-\beta\left[-\frac{1}{2}\,\mathrm{cosh}\left(\frac{\phi_{-}}{2}\right)\,F_{-}+\frac{i}{4}\,\mathrm{sinh}\left(\frac{\phi_{-}}{2}\right)\,\psi_{-}\bar{\psi}_{-}\right],\\
b_{2}&=&\beta\,\mathrm{senh}\left(\frac{\phi_{-}}{2}\right),\\
\partial_{z}b_{2}&=&\frac{2i}{\beta}\left[-\frac{1}{2}\,\mathrm{cosh}\left(\frac{\phi_{+}}{2}\right)\,F_{+}+\frac{i}{4}\,\mathrm{sinh}\left(\frac{\phi_{+}}{2}\right)\,\psi_{+}\bar{\psi}_{+}\right],\\
f_{2}&=&-\frac{\beta}{2}\,\mathrm{cosh}\left(\frac{\phi_{-}}{2}\right)\,\bar{\psi}_{-}=-\frac{1}{\beta}\,\mathrm{cosh}\left(\frac{\phi_{+}}{2}\right)\,\psi_{+},\\
F_{-}&=&-\frac{4i}{\beta}\left[\frac{f_{1}}{2}\,\mathrm{sinh}\left(\frac{\phi_{+}}{2}\right)\psi_{+}-b_{2}\,\mathrm{cosh}\left(\frac{\phi_{+}}{2}\right)\right],\\
F_{+}&=&-2\beta\left[\frac{f_{1}}{2}\,\mathrm{sinh}\left(\frac{\phi_{-}}{2}\right)\bar{\psi}_{-}-b_{1}\,\mathrm{cosh}\left(\frac{\phi_{-}}{2}\right)\right],\\
\partial_{z}\phi_{-}&=&\frac{4i}{\beta}\left[\frac{f_{1}}{2}\,\mathrm{sinh}\left(\frac{\phi_{+}}{2}\right)\bar{\psi}_{+}-b_{1}\,\mathrm{cosh}\left(\frac{\phi_{+}}{2}\right)\right],\\
\partial_{\bar{z}}\phi_{+}&=&-2\beta\left[\frac{if_{1}}{2}\,\mathrm{sinh}\left(\frac{\phi_{-}}{2}\right)\psi_{-}-b_{2}\,\mathrm{cosh}\left(\frac{\phi_{-}}{2}\right)\right],\\
i\partial_{z}\psi_{-}&=&\frac{4i}{\beta}\left[\frac{f_{1}}{2}\,\mathrm{sinh}\left(\frac{\phi_{+}}{2}\right)\,F_{+}-\frac{if_{1}}{4}\,\mathrm{cosh}\left(\frac{\phi_{+}}{2}\right)\psi_{+}\bar{\psi}_{+}\right]\nonumber\\
&&-\frac{4i}{\beta}\left[\frac{ib_{1}}{2}\,\mathrm{sinh}\left(\frac{\phi_{+}}{2}\right)\psi_{+}-\frac{b_{2}}{2}\,\mathrm{sinh}\left(\frac{\phi_{+}}{2}\right)\,\bar{\psi}_{+}\right]\nonumber\\
&&-\frac{4i}{\beta}\,f_{2}\,\mathrm{cosh}\left(\frac{\phi_{+}}{2}\right),\\
\partial_{\bar{z}}\bar{\psi}_{+}&=&-2\beta\left[-\frac{f_{1}}{2}\,\mathrm{sinh}\left(\frac{\phi_{-}}{2}\right)\,F_{-}+\frac{if_{1}}{4}\,\mathrm{cosh}\left(\frac{\phi_{-}}{2}\right)\psi_{-}\bar{\psi}_{-}\right]\nonumber\\
&&-2\beta\left[\frac{ib_{1}}{2}\,\mathrm{sinh}\left(\frac{\phi_{-}}{2}\right)\psi_{-}-\frac{b_{2}}{2}\,\mathrm{sinh}\left(\frac{\phi_{-}}{2}\right)\,\bar{\psi}_{-}\right]\nonumber\\
&&-2\beta\,f_{2}\,\mathrm{cosh}\left(\frac{\phi_{-}}{2}\right).\\
\er
Eliminating the unphysical fields, we find 
\begin{eqnarray}
\partial_{z}\phi_{-}&=&-\frac{1}{2}\,\mathrm{tanh}\left(\frac{\phi_{+}}{2}\right)\,
\bar{\psi}_{-}\bar{\psi}_{+}+\frac{4}{\beta^{2}}\mathrm{sinh}\,\phi_{+},\\[12pt]
\partial_{z}\bar{\psi}_{-}&=&\frac{1}{2}\bar{\psi}_{-}\,
\mathrm{tanh}\left(\frac{\phi_{+}}{2}\right)\partial_{z}\,\phi_{+}+\frac{4}{\beta^{2}}\,
\mathrm{cosh}^{2}\left(\frac{\phi_{+}}{2}\right)\,\bar{\psi}_{+}\\[12pt]
\partial_{\bar z}\phi_{-}&=&\frac{\beta^{2}}{8}\,
\frac{\mathrm{tanh}\left(\phi_{-}/2\right)}{\mathrm{cosh}^{2}\left(\phi_{+}/2\right)}\,
\bar{\psi}_{-}\partial_{t}\bar{\psi}_{-}+\beta^{2}\,\mathrm{sinh}\,\phi_{-},\\[12pt]
\partial_{\bar z}\bar{\psi}_{+}&=&
\mathrm{tanh}\left(\frac{\phi_{+}}{2}\right)\mathrm{tanh}\left(\frac{\phi_{-}}{2}\right)\,
\partial_{\bar z}\bar{\psi}_{-}\nonumber\\
&-&\frac{\beta^{2}}{2}\,\mathrm{tanh}^{2}\left(\frac{\phi_{+}}{2}\right)\mathrm{tanh}\left(\frac{\phi_{-}}{2}\right)\mathrm{sinh}\,
\phi_{-}\,\bar{\psi}_{-}\nonumber\\
&+&\beta^{2}\,\mathrm{cosh}^{2}\left(\frac{\phi_{-}}{2}\right)\,\bar{\psi}_{-}.
\end{eqnarray}

\end{document}